\documentclass[aps,amsmath,amssymb,twocolumn,10pt,superscriptaddress]{revtex4-1}
\usepackage{graphicx}  
\usepackage[dvipsnames]{xcolor}
\usepackage[colorlinks=true,linkcolor=blue,citecolor=red,urlcolor=magenta]{hyperref}
\usepackage{utopia}

\begin{document}
 
\title{Lissajous dynamics of a quantum particle in a tilted two-dimensional discrete lattice}
\author{Grzegorz Jaczewski}
\affiliation{Faculty of Physics, University of Warsaw, ul. Pasteura 5, PL-02093 Warsaw, Poland}
\affiliation{Institute of Physics, Polish Academy of Sciences, Aleja Lotnik\'ow 32/46, PL-02668 Warsaw, Poland} 
\author{Tomasz Sowi\'nski} 
\affiliation{Institute of Physics, Polish Academy of Sciences, Aleja Lotnik\'ow 32/46, PL-02668 Warsaw, Poland} 
\date{\today} 

\begin{abstract}
The quantum dynamics of a single particle in a discrete two-dimensional tilted lattice is analyzed from the perspective of the classical-quantum correspondence. Utilizing the fact that tilting the lattice results in oscillatory dynamics, we show how the parameters of the lattice and the initial state of the particle can be tuned so that during evolution the probability distribution does not change its shape while its center follows the trajectory known in classical mechanics as Lissajous curves.
\end{abstract}
 
\maketitle

\section{Introduction}
The question of the relationship between quantum dynamics and its classical counterpart has been analyzed in many different ways since the birth of quantum mechanics. In the simplest cases of systems described by quadratic Hamiltonians (both time-dependent and time-independent) in any number of dimensions, {\it e.g.}, such as the harmonic oscillator, it can be shown straightforwardly that the evolution of the expectation values of certain operators is exactly the same as the evolution of the classical counterparts~\cite{1926Schrodinger,1954Senitzky,1982Hartley,1999Andrews}. This observation has been elegantly generalized by Ehrenfest to all mechanical systems, with his theorem showing exactly where deviations of the quantum mechanical description from the classical one occur~\cite{1927Ehrenfest}. These approaches have allowed also the discovery of approximate solutions of the Schr\"odinger equation describing a particle moving in an arbitrary electromagnetic field and conclusively to introduce so-called trajectory‐coherent states, {\it i.e.}, specific-shape wave packets following trajectories determined by the classical equation of motion~\cite{1983Bagrov}. 

In our work, we explore a slightly different aspect of the correspondence between classical and quantum dynamics. We base this on the well-known observation that a quantum particle moving in a periodic potential subjected to the additional influence of a constant force performs specific oscillations, the so-called Bloch oscillations~\cite{1929BlochZPhysik,1934Zener}. Extending this, we perform a detailed analysis of the two-dimensional dynamics of a localized wave packet in a discrete two-dimensional lattice and show that the parameters of the system can be tuned so that the trajectory traced by the packet has the shape of the Lissajous curves known from classical physics, {\it i.e.}, trajectories drawn by a classical oscillating particle in two perpendicular directions. In this way, we organise previously known results~\cite{2003Kolovsky,2004Witthaut,2005Mossmann,2006Breid,2017Carrillo}
 and give a clear classification of possible curves and the parameters realizing them. In contrast to previous attempts, we consider dynamics in a purely discrete system. In this way, we extract the simplest generic system manifesting quantum Lissajous dynamics.
 
It is worth mentioning that we consider a scenario with a coin-free walker, {\it i.e.}, a quantum particle hopping to neighboring sites with equal probabilities independent of its internal quantum state. From this perspective, our system is essentially different from generic discrete quantum walk systems studied extensively in the literature in many different contexts in which spatial dynamics is substantially entangled with the internal degree of freedom of the walker\cite{1993Aharonov,2004Wojcik,2010Kitagawa,2010WitthautPRA,2013Cedzich,2015PhysRevA,2018Perez,2020ArnaultPRA,2024Wojcik}. 
 
\section{The model}
In our work, we consider the simplest possible scenario of a quantum particle moving in a two-dimensional square lattice, additionally subjected to the external constant force. Assuming that a vector $|\mathsf{x},\mathsf{y}\rangle$ describes a quantum state of a particle occupying lattice site with coordinates $(\mathsf{x},\mathsf{y})$ (where $\mathsf{x}$ and $\mathsf{y}$ are integers) and the tunneling is allowed only to the neighboring lattice sites, one can write the Hamiltonian of the system as
\begin{equation} \label{Hamiltonian}
    \hat{H}=\hat{H}_0 + \hat{H}_F,
\end{equation}
where
\begin{align}
\hat{H}_0 = &-\sum_{\mathsf{x},\mathsf{y}}J_x(|\mathsf{x}-1,\mathsf{y}\rangle+|\mathsf{x}+1,\mathsf{y}\rangle)\langle \mathsf{x},\mathsf{y}| \nonumber \\
&-\sum_{\mathsf{x}y}J_y(|\mathsf{x},\mathsf{y}-1\rangle+|\mathsf{x},\mathsf{y}+1\rangle)\langle \mathsf{x},\mathsf{y}|
\end{align}
describes the dynamics in the homogenous lattice unaffected by external force, while
\begin{equation}
\hat{H}_F = \sum_{\mathsf{x},\mathsf{y}}\left(\mathsf{x} F_x+y F_y\right)|{\mathsf{x},\mathsf{y}}\rangle\langle{\mathsf{x},\mathsf{y}}|
\end{equation}
introduces external linear potential energy shift with slopes $F_x$ and $F_y$. These parameters have a direct interpretation of mutually perpendicular components of the external constant force acting on the particle (for example external electric field). Although in general, tunneling in perpendicular directions can be tuned independently, in the following we will set both amplitudes equal $J_x=J_y=J$. 

At any moment $t$ the quantum state of moving particle $|\Psi(t)\rangle$ can be decomposed into basis states $\{|\mathsf{x},\mathsf{y}\rangle\}$ as $|\Psi(t)\rangle=\sum_{\mathsf{x}\mathsf{y}}\psi_{\mathsf{x}\mathsf{y}}(t)|\mathsf{x},\mathsf{y}\rangle$, where $\psi_{\mathsf{x}\mathsf{y}}(t)$ has a natural interpretation of the probability amplitude of finding a particle at lattice site $(\mathsf{x},\mathsf{y})$. In this framework, evolution is provided by a set of dynamical equations derived directly from the Schr\"odinger equation 
\begin{equation}
i\hbar \frac{\mathrm{d}}{\mathrm{d}t}\psi_{\mathsf{x}\mathsf{y}}(t) = \sum_{\mathsf{x}'\mathsf{y}'} \langle \mathsf{x},\mathsf{y}|\hat{H}|\mathsf{x}'\mathsf{y}'\rangle\psi_{\mathsf{x}',\mathsf{y}'}(t).
\end{equation}

To form a direct bridge with a classical concept of the Lissajous curves, in the following, we consider situations in which quantum particle is initially localized in space and momentum domains. Formally it means that variances of expectation values of corresponding operators are small enough and they can be used to introduce the concept of trajectory of a quantum particle, that later can be compared with its classical counterpart. In our approach we consider the simplest possible initial state fulfilling these requirements in the form of Gaussian wave-packet
\begin{equation} \label{gaussian}
\psi_{\mathsf{x}\mathsf{y}}(0)={\cal N}^2\mathrm{exp}\left[-\frac{(\mathsf{x}-{\cal X})^2+(\mathsf{y}-{\cal Y})^2}{4\sigma^2}+i(\mathsf{x}{\cal P}_x + \mathsf{y}{\cal P}_y)\right],
\end{equation}
where ${\cal X},{\cal Y}$ and ${\cal P}_x,{\cal P}_y$ are average position and momentum 
of the wave-packet, $\sigma$ defines its spatial width, and ${\cal N}$ is numerical constant guaranteing appropriate normalization, $\sum_{\mathsf{x}\mathsf{y}}|\psi_{\mathsf{x}\mathsf{y}}(0)|^2=1$. One of our aims is to determine widths $\sigma$ for which classical-quantum similarity of the dynamics is clearly visible. 

\section{One-dimensional dynamics}
\begin{figure}
\includegraphics[width=1\linewidth]{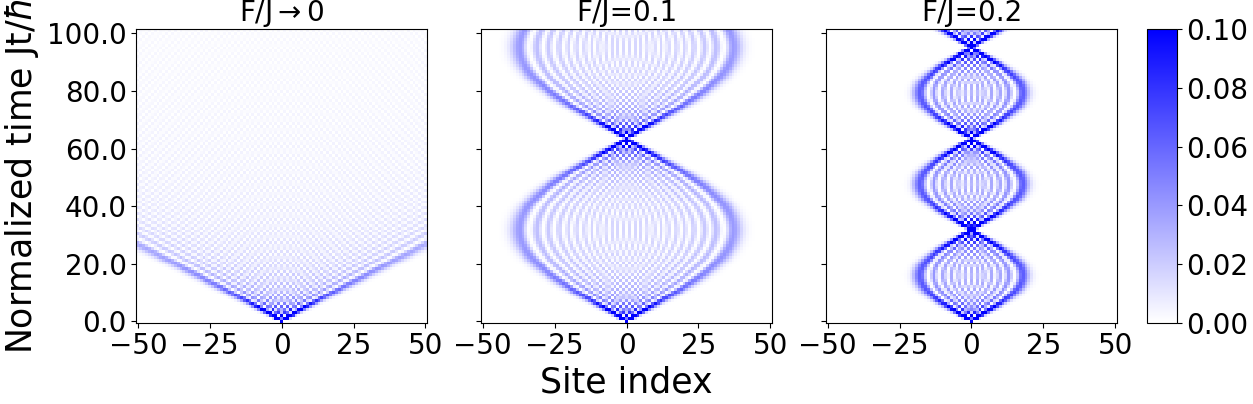}
\caption{Time evolution of the density distribution of the particle initially localized in selected lattice site ($\mathsf{x}_0=0$) for different intensities of the external force $F$. Periodic oscillations of the density around the initial position are clearly visible. In the limit of vanishing force, a diffusive solution \eqref{DiffusiveSolution} is restored. In all plots, time is measured in natural unit $\hbar/J$.}
    \label{Fig1}
   \end{figure}

Before we start to analyze the dynamics in a two-dimensional scenario let us first recall known results in the one-dimensional case. This is simply the limiting case of the original problem \eqref{Hamiltonian} when $J_y$ is set to $0$. In this case, the Hamiltonian reduces simply to the following form:
\begin{equation} \label{Ham1D}
\hat{H} = -\sum_{\mathsf{x}}\big(J|\mathsf{x}-1\rangle+J|\mathsf{x}+1\rangle- \mathsf{x} F|\mathsf{x}\rangle\big)\langle \mathsf{x}|,
\end{equation}
where $J$ and $F$ are tunneling and slope along the chain, respectively. It is known that the Hamiltonian \eqref{Ham1D} is diagonal in the following basis of eigenstates enumerated with integer $n$ \cite{old,Holthaus}
\begin{equation}\label{eq:osob}
    |n\rangle=\sum_{\mathsf{x}}\mathcal{J}_{\mathsf{x}-n}\left(\frac{2J}{F}\right)|\mathsf{x}\rangle,
\end{equation}
where ${\cal J}$ is the Bessel function of the first kind. Corresponding eigenvalues are expressed simply as $\lambda_{n}=n F$. Thus, if initially, the particle's wave function has a form $|\Psi(0)\rangle=\sum_{\mathsf{x}}\psi_{\mathsf{x}}(0)|\mathsf{x}\rangle$,
one straightforwardly finds the wave function at an arbitrary moment
\begin{equation}\label{WaveFunction}
|\Psi(t)\rangle=\sum_{n,\mathsf{x},\mathsf{x}'}\psi_{\mathsf{x}'}(0)\mathcal{J}_{\mathsf{x}'-n}\!\left(\frac{2J}{F}\right)\mathcal{J}_{\mathsf{x}-n}\!\left(\frac{2J}{F}\right)\mathrm{e}^{-inFt/\hbar}|\mathsf{x}\rangle.
\end{equation}
This expression can be simplified further by utilizing the generalized sum rule for Bessel functions
\begin{equation} \label{BesselIdentity}
    \sum_n\mathcal{J}_n(z)\mathcal{J}_{\mathsf{x}+n}(z)e^{in\phi}=\mathcal{J}_{\mathsf{x}}\!\left[2z\sin\left(\frac{\phi}{2}\right)\right]e^{i\mathsf{x}(\pi-\phi)/2}.
\end{equation}
After applying this identity and appropriate shifting of sums one finds temporal probability amplitudes of finding a particle at individual lattice sites
\begin{align} \label{SolutionGeneral}
    \psi_\mathsf{x}(t)&=\langle \mathsf{x}|\Psi(t)\rangle = \\
    &\sum_{\mathsf{x}'}\psi_{\mathsf{x}'}(0)\mathcal{J}_{\mathsf{x}-\mathsf{x}'}\!\left[\frac{4J}{F}\sin\left(\frac{Ft}{2\hbar}\right)\right]\mathrm{e}^{\frac{i}{2}\left[\pi(\mathsf{x}-\mathsf{x}')-\frac{F(\mathsf{x}+\mathsf{x}')t}{\hbar}\right]}. \nonumber
\end{align} 
In the case of the particle initially prepared in the Gaussian state \eqref{gaussian} (reduced to one dimension along $X$ axis) we can explicitly write the final expression
\begin{align}\label{eq:psij}
    \psi_\mathsf{x}(t)&=\langle \mathsf{x}|\Psi(t)\rangle = \nonumber \\
    &{\cal N}\sum_{\mathsf{x}'}\mathcal{J}_{\mathsf{x}-\mathsf{x}'}\!\left[\frac{4J}{F}\sin\left(\frac{Ft}{2\hbar}\right)\right] \times \\ 
    &\exp\left[-\frac{(\mathsf{x}'-{\cal X})^2}{4\sigma^2}+i\mathsf{x}'{\cal P}+\frac{i\pi(\mathsf{x}-\mathsf{x}')}{2}-\frac{i(\mathsf{x}+\mathsf{x}')Ft}{2\hbar}\right]. \nonumber
\end{align}
Let us now consider several interesting limiting cases of this result. First, we note that in the limit of vanishing external force $F\rightarrow 0$, one straightforwardly restores a well-known formula for the evolution of probability amplitudes for a quantum diffusion of Gaussian wave packet~\cite{Hartmann}
\begin{multline}\label{eq:psimpor}
\psi_{\mathsf{x}}(t)=
{\cal N}\sum_{\mathsf{x}'}\mathcal{J}_{\mathsf{x}-\mathsf{x'}}\left(\frac{2Jt}{\hbar}\right) \\
\times\exp\left[-\frac{(\mathsf{x}'-{\cal X})^2}{4\sigma^2}+i\mathsf{x}'{\cal P}+\frac{i\pi(\mathsf{x}-\mathsf{x}')}{2}\right].
\end{multline}
Another interesting case is obtained when the external force is present and the particle is initially exactly localized at one of the lattice sites $\mathsf{x}_0$, {\it i.e.}, $\psi_x(0)=\delta_{\mathsf{x},\mathsf{x}_0}$. Then the general solution \eqref{SolutionGeneral} is significantly simplified and reduces to
\begin{equation}
    \psi_\mathsf{x}(t)=\mathcal{J}_{\mathsf{x}-\mathsf{x}_0}\!\left[\frac{4J}{F}\sin\left(\frac{Ft}{2\hbar}\right)\right] \mathrm{e}^{\frac{i}{2}\left[\pi(\mathsf{x}-\mathsf{x}_0)-\frac{F(\mathsf{x}+\mathsf{x}_0)t}{\hbar}\right]}.
\end{equation}
This solution shows that in the case of an initially localized particle, its spatial distribution periodically changes in time and alternately expands and contracts around the initial position $\mathsf{x}_0$. This behavior is displayed in Fig.~\ref{Fig1} for different $F$. Again, in the limit of vanishing force $F/J\rightarrow 0$, we restore a well-known diffusive solution manifesting characteristic interference pattern
\begin{equation} \label{DiffusiveSolution}
    \psi_\mathsf{x}(t)=\mathcal{J}_{\mathsf{x}-\mathsf{x}_0}\left(\frac{2Jt}{\hbar}\right)\mathrm{e}^{i\pi(\mathsf{x}-\mathsf{x}_0)/2}.
\end{equation} 

Finally, let us examine the most important for further analysis case of particle initially significantly spread, {\it i.e.}, when the spatial width of the initial wave packet is large when compared to the distance between lattice sites, $\sigma\gg 1$. Then, it is very convenient to express the general solution \eqref{SolutionGeneral} in terms of the Fourier transform of the initial wave function 
$
\tilde\psi(k) = \frac{1}{2\pi}\sum_\mathsf{x} \psi_\mathsf{x}(0)\mathrm{e}^{-ik\mathsf{x}}
$
since in the momentum domain, the wave function is well-localised. After substituting an inverse relation
$
\psi_\mathsf{x}(0) = \int \mathrm{d}k\,\tilde\psi(k)\mathrm{e}^{ik\mathsf{x}}
$
into general solution \eqref{SolutionGeneral} and performing some straightforward algebraic transformations one finds
\begin{multline} \label{SolutionGeneral2}
\psi_\mathsf{x}(t) = \int\!\mathrm{d}k\,\tilde\psi(k) \mathrm{e}^{i\mathsf{x}(k-Ft/\hbar)}\\ \times \mathrm{exp}\left[\frac{4iJ}{F}\sin\left(\frac{Ft}{2\hbar}\right)\cos\left(k-\frac{Ft}{2\hbar}\right)\right].
\end{multline}
Let us emphasize that expression \eqref{SolutionGeneral2} is exactly equivalent to relation \eqref{SolutionGeneral} since any approximation has not been introduced so far. Now, we exploit the fact that for large $\sigma$ amplitude $\psi_k(0)$ is well-localized around initial momentum ${\cal P}$ and therefore we expand the last term in the difference $(k-{\cal P})$ and keep  only constant and linear term 
\begin{multline}\label{Expansion}
\cos\left(k-\frac{Ft}{2\hbar}\right) \approx \\ \cos\left({\cal P}-\frac{Ft}{2\hbar}\right)-\sin\left({\cal P}-\frac{Ft}{2\hbar}\right)(k-{\cal P}).
\end{multline}
\begin{figure}
\includegraphics[width=\linewidth]{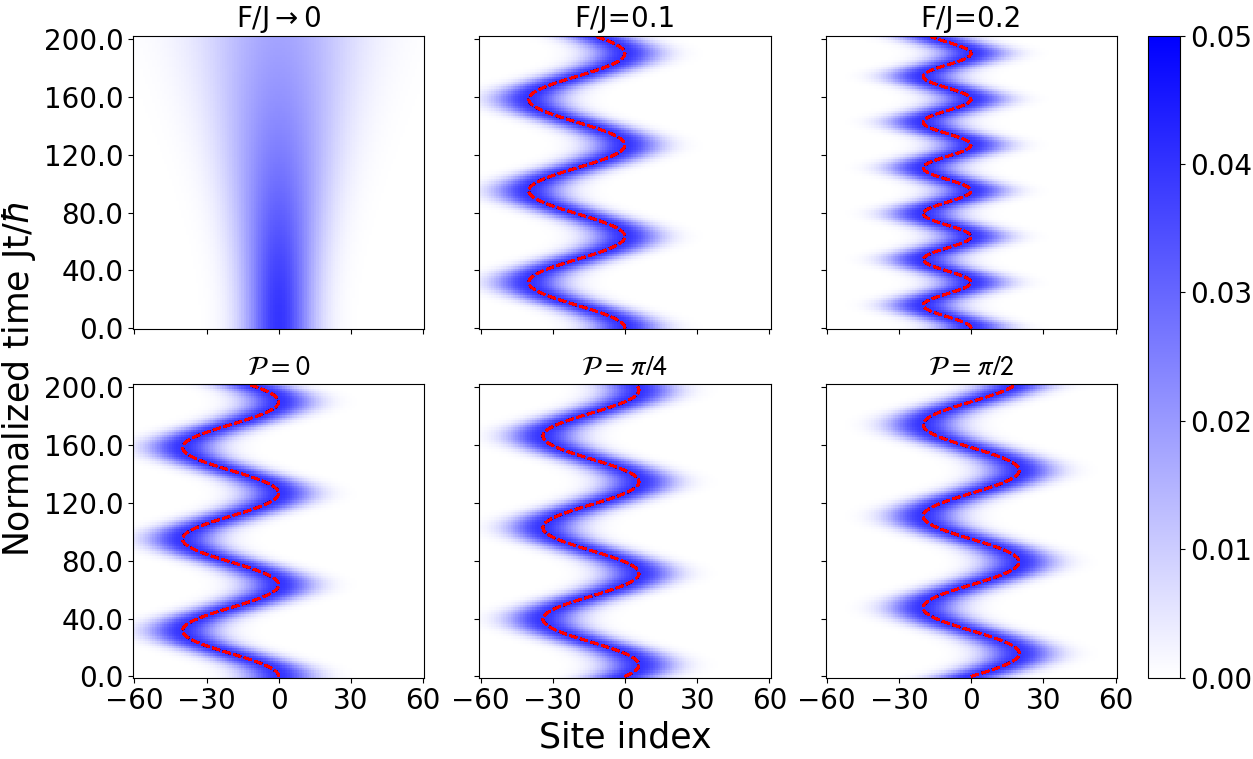}
\caption{Time evolution of the density distribution of the particle initially prepared in a wide Gaussian state, $\sigma=10$, centered around $\mathsf{x}_0=0$. (top row) Results for different intensities of external force $F/J$ and initial momentum ${\cal P}=0$. (bottom row) Results for different initial momenta ${\cal P}$ and particular intensity of external force $F/J=0.1$. In all plots, red dashed lines represent the corresponding approximate solutions \eqref{DeltaTime} and time is measured in natural unit $\hbar/J$.}
\label{Fig2}
\end{figure} 
Then, after using explicit expression for the initial wave function and performing integration, one finds the time dependence of probability amplitudes in the position representation
\begin{equation} \label{1dGaussianMotion}
\psi_{\mathsf{x}}(t)={\cal N}\mathrm{exp}\left[-\frac{(\mathsf{x}-\Delta(t))^2}{4\sigma^2}+i\mathsf{x}\Gamma(t)+i\Phi(t)\right],
\end{equation}
where
\begin{subequations} \label{Center1d}
\begin{align}
\Delta(t) &={\cal X}+\frac{2J}{F}\left[\cos(Ft/\hbar-{\cal P})-\cos({\cal P})\right], \label{DeltaTime}\\
\Gamma(t)&={\cal P}-Ft/\hbar, \\
\Phi(t)&=\frac{2J}{F}[\sin(Ft/\hbar-\cal P)+\sin(\cal P).
\end{align}
\end{subequations}
This result clearly shows that a sufficiently wide Gaussian wave packet preserves its spatial shape during the evolution. At the same time, its center harmonically oscillates with frequency proportional to the force $F$ and amplitude proportional to the ratio $J/F$. This counterintuitive behavior of a quantum wave packet in a tilted periodic potential is known as Bloch oscillations~\cite{1929BlochZPhysik,1934Zener} and was not experimentally confirmed until 1992  \cite{bloch2,bloch3}. The exact evolution of a wave packet with $\sigma=10$ for various tilting and initial momentum ${\cal P}$ are presented in Fig.~\ref{Fig2}. For clarity, in all plots, we mark with a dashed red line a trace of the center of the wavepacket as predicted by approximation \eqref{DeltaTime}. Agreement between the two results is clearly visible. 

\begin{figure}
\includegraphics[width=\linewidth]{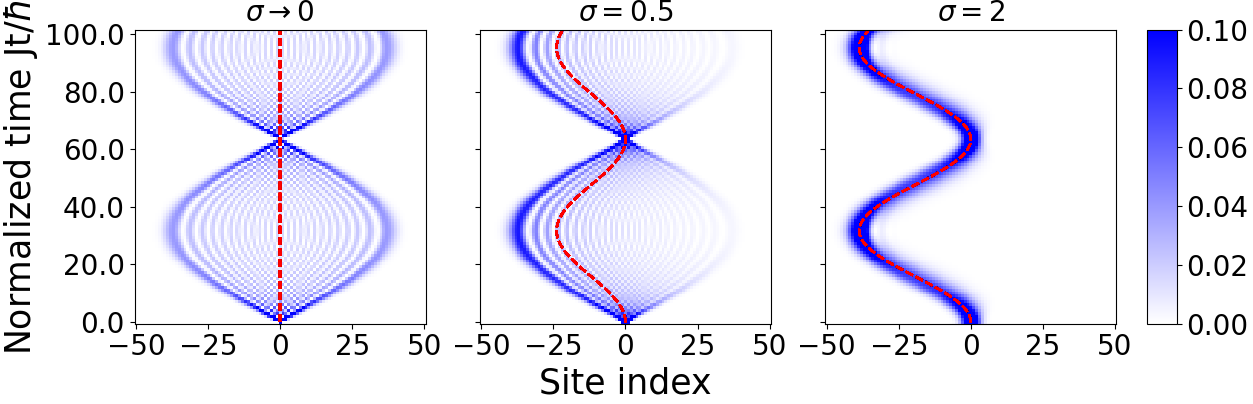}
\caption{
Time evolution of the density distribution of the particle initially prepared in Gaussian states of different widths subjected to the external force of fixed intensity $F/J=0.1$. The transition from the breathing dynamics ($\sigma\rightarrow 0$) to coherent oscillations of the center of the packet ($\sigma\gg 1)$ is clearly visible. All results obtained for initial parameters $\mathsf{x}_0=0$ and ${\cal P}=0$. In all plots, red dashed lines represent the position of the center of the distribution as predicted exactly by \eqref{ExpPosition}. Time is measured in natural unit $\hbar/J$.}
\label{Fig3}
\end{figure} 
For completeness of the discussion, in Fig.~\ref{Fig3} we also show how the behavior of the wave function changes from breathing mode to oscillations when one varies the width of the initial wave packet $\sigma$ from very small to very large. This non-obvious behavior of the system can be better understood when the expectation value of the position operator is considered. After taking the general solution for Gaussian initial condition \eqref{eq:psij} one can explicitly derive this quantity's time dependence. It turns out that it manifests specific oscillations
\begin{align}
 \langle \mathsf{x}(t)\rangle&= \sum_\mathsf{x}\mathsf{x}\,|\psi_\mathsf{x}(t)|^2 \nonumber \\
 &={\cal X}+{\cal A}({\cal X},\sigma)\left[\cos\left(\frac{Ft}{\hbar}-{\cal P}\right)-\cos\left({\cal P}\right)\right].\label{ExpPosition}
\end{align}
The amplitude of these oscillations ${\cal A}({\cal X},\sigma)$ non-trivially depends on the initial width $\sigma$ and initial position of the wave packet ${\cal X}$
\begin{subequations}
\begin{align}
{\cal A}({\cal X},\sigma)=\frac{2J}{F}\,\frac{\sqrt{2\pi}\sigma}{\mathrm{e}^{1/8\sigma^2}}\,\frac{\vartheta_3\left(\frac{\pi}{2}+\pi{\cal X},e^{-2\pi^2\sigma^2}\right)}{\vartheta_3\left(0,e^{-1/2\sigma^2}\right)}
\end{align}
where $\vartheta_3$ is the elliptic theta function defined as
\begin{equation}
\vartheta_3(x,y)=1+2\sum_{n=1}^\infty y^{n^2}\cos(2nx).
\end{equation}
\end{subequations}
Directly from the properties of the function $\vartheta_3$ one can show that in the case of a perfectly localized particle, $\sigma\rightarrow 0$, the amplitude ${\cal A}$ vanishes. Therefore, the center of the distribution has a fixed position equal to the initial position ${\cal X}$, which recovers the numerical result presented in the left panel of Fig.~\ref{Fig3}. Contrary, for finite widths $\sigma$, the distribution oscillates around the position determined by the initial parameters of the wave packet, ${\cal X}$ and ${\cal P}$. The amplitude of these oscillations increases monotonically with $\sigma$, and in the limit $\sigma\rightarrow\infty$ it saturates at $2J/F$, restoring the result \eqref{DeltaTime}. Time evolution of the expectation value $\langle\mathsf{x}(t)\rangle$ is visualized in Fig.~\ref{Fig3} with red dashed lines. 

A very similar analysis can be performed for the evolution of the variance of the density distribution
\begin{equation}
s^2(t)=\sum_\mathsf{x}\mathsf{x}^2\,|\psi_\mathsf{x}(t)|^2-\langle \mathsf{x}(t)\rangle^2.
\end{equation}
After some rather more arduous but still straightforward calculations, it can be shown that
\begin{align}
s^2(t) = s^2(0) + \left[\frac{4J}{F}\sin\left(\frac{Ft}{2\hbar}\right)\right]^2\left[\frac{1}{2}-{\cal S}(\sigma,t)\right],
\end{align}
where
\begin{subequations}
\begin{align}
s^2(0) &=\mathrm{e}^{-1/2\sigma^2}\frac{\vartheta_3'(0,\mathrm{e}^{-1/2\sigma^2})}{\vartheta_3(0,\mathrm{e}^{-1/2\sigma^2})}, \\
{\cal S}(\sigma,t)&= \\ 
&\sqrt{\frac{\pi}{2}}\sigma\mathrm{e}^{-1/2\sigma^2}\frac{\vartheta_3(0,\mathrm{e}^{-2\pi^2\sigma^2})}{\vartheta_3(0,\mathrm{e}^{-1/2\sigma^2})}\cos\left(\frac{Ft}{\hbar}-2{\cal P}\right) \nonumber \\
&+2\pi\sigma^2\mathrm{e}^{-1/4\sigma^2}\frac{\vartheta_3(\pi/2,\mathrm{e}^{-2\pi^2\sigma^2})}{\vartheta_3^2(0,\mathrm{e}^{-1/2\sigma^2})}\sin^2\left(\frac{Ft}{2\hbar}-{\cal P}\right). \nonumber
\end{align}
\end{subequations}
For convenience, we introduced here also a shortened notation for the derivative of the elliptic function along the second argument, $\vartheta_3'(u,q)=\mathrm{d}\vartheta_3(u,q)/\mathrm{d}q$.

With this result, we can easily understand the transition between different behaviors displayed in Fig.~\ref{Fig3}. First, in the limit of perfectly localized initial wavepacket, $\sigma\rightarrow 0$, one finds that the function ${\cal S}(\sigma,t)$ vanishes for any time $t$ while $s^2(0)\rightarrow 0$. Thus, in this limit, the variance of the distribution simply oscillates as
\begin{align}
s^2(t)=\frac{8J^2}{F^2}\sin^2\left(\frac{Ft}{2\hbar}\right).
\end{align}
Contrary, in the limit of very wide initial distribution, $\sigma\rightarrow\infty$, we 
surprisingly find that the function ${\cal S}(\sigma,t)$ again is time independent and equal to $1/2$. Moreover, in this limit, initial variance $s^2(0)\rightarrow\sigma^2$. In consequence, the variance of the distribution becomes time-independent and equal to the variance of the initial Gaussian state $\sigma^2$. The evolution of the distribution for different $\sigma$ displayed in Fig.~\ref{Fig3} nicely reflects the transition between these two limits.

\section{Dynamics in two dimensions}
After an extended discussion of the dynamics in one dimension, let us now switch to the initial problem of our work. The general aim is to adjust physical parameters of the system, {\it i.e.}, parameters of the Hamiltonian as well as the particle's initial state, to make the wave packet move along trajectories being counterparts of the classical Lissajous figures. To make a first step in this direction, let us recall one-dimensional result \eqref{1dGaussianMotion} explaining why a sufficiently broad wave packet moves in the lattice in an oscillatory manner without changing its shape. In full analogy, due to the separability of motions in perpendicular directions guaranteed by the Hamiltonian \eqref{Hamiltonian}, in a two-dimensional case, the resulting wave function of a sufficiently wide Gaussian state has a form
\begin{multline}
\psi_{\mathsf{xy}}(t)={\cal N}\mathrm{exp}\left[-\frac{(\mathsf{x}-\Delta_x(t))^2+(\mathsf{y}-\Delta_y(t))^2}{4\sigma^2}\right] \\ \times\mathrm{e}^{i\mathsf{x}\Gamma_x(t)+i\mathsf{y}\Gamma_y(t)+i\Phi(t)},
\end{multline}
where
\begin{subequations}
\begin{align}\label{eq:uklad}
\Delta_x(t) &={\cal X}+\frac{2J}{F_x}\left[\cos\left(F_x t/\hbar-{\cal P}_x\right)-\cos({\cal P}_x)\right],\\
\Delta_y(t)&={\cal Y}+\frac{2J}{F_y}\left[\cos\left(F_y t/\hbar-{\cal P}_y\right)-\cos({\cal P}_y)\right], \\
\Gamma_x(t) &={\cal P}_x-F_xt/\hbar, \\
\Gamma_y(t) &={\cal P}_y-F_yt/\hbar, \\
\Phi(t) &=\frac{2J}{F_x}[\sin(F_xt-{\cal P}_x)+\sin({\cal P}_x)] \nonumber \\
&+\frac{2J}{F_y}[\sin(F_yt-{\cal P}_y)+\sin({\cal P}_y)].
\end{align}
\end{subequations}
\begin{figure} 
\includegraphics[width=\linewidth]{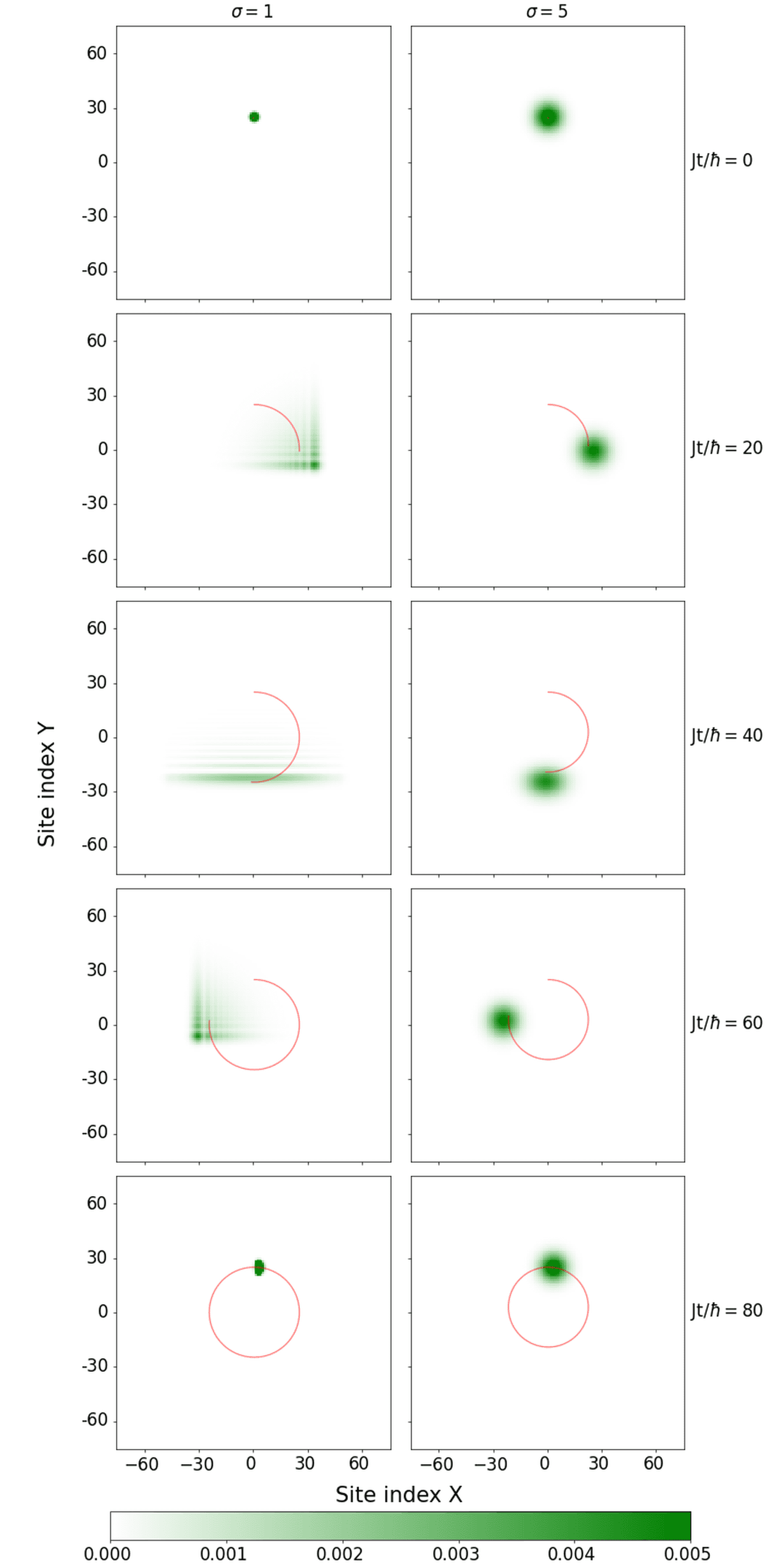}
\caption{Dynamics of the wave packet in a tilted lattice tuned to a circular Lissajous curve ($2J/F_x=2J/F_y=25$). The left (right) column corresponds to the case of the particle initially prepared in a narrow (wide) Gaussian state. In both cases, initial momentum ${\cal P}_x=\pi/2$. Consecutive rows correspond to different moments of the evolution. Note that the classical trajectory is restored for sufficiently large widths, {\it i.e.}, when the temporal shape of the distribution does not change significantly. Corresponding movies presenting a whole evolution are accessible online~\cite{2024JaczewskiSowinskiZenodo}.}
    \label{Fig4}
\end{figure} 
By setting the initial position and momenta as 
\begin{equation}\label{eq:pocz}
{\cal X}=\frac{2J}{F_x}\cos({\cal P}_x), \qquad
{\cal Y}=\frac{2J}{F_y}, \qquad
{\cal P}_y=0
\end{equation}
we find that the center of the wave packet moves along the celebrated Lissajous curve parametrically described by
\begin{subequations}
\begin{align}
\Delta_x(t) &=A\cos(\Omega_x t + \varphi), \\
\Delta_y(t) &=B\cos(\Omega_y t),
\end{align}
\end{subequations}
where $A=2J/F_x$, $B=2J/F_y$, $\Omega_x=F_x/\hbar$, $\Omega_y=F_y/\hbar$, and $\varphi=-{\cal P}_x$. Of course, the above reasoning is valid as long as the wave packet remains localized during a whole evolution. To clarify this, let us focus on the simplest Lissajous curve -- a circle obtained for $\Omega_x=\Omega_y$, $A=B$, and $\varphi=\pi/2$. In Fig.~\ref{Fig4} we show the time evolution of the density distribution for parameters tailored to this scenario, {\it i.e.}, $2J/F_x=2J/F_y=25$, ${\cal P}_x=\pi/2$. The right panel presents the evolution for a sufficiently wide packet with $\sigma=5$, and the distribution nicely follows the corresponding Lissajous curve (red line). On the contrary, in the case of the too-narrow wave packet with $\sigma=1$ (left panel), the density significantly changes its shape during evolution, and the reasoning clearly breaks up. For convenience, movies showing these evolutions are accessible online~\cite{2024JaczewskiSowinskiZenodo}.

\begin{figure}   
\includegraphics[width=1.\linewidth]{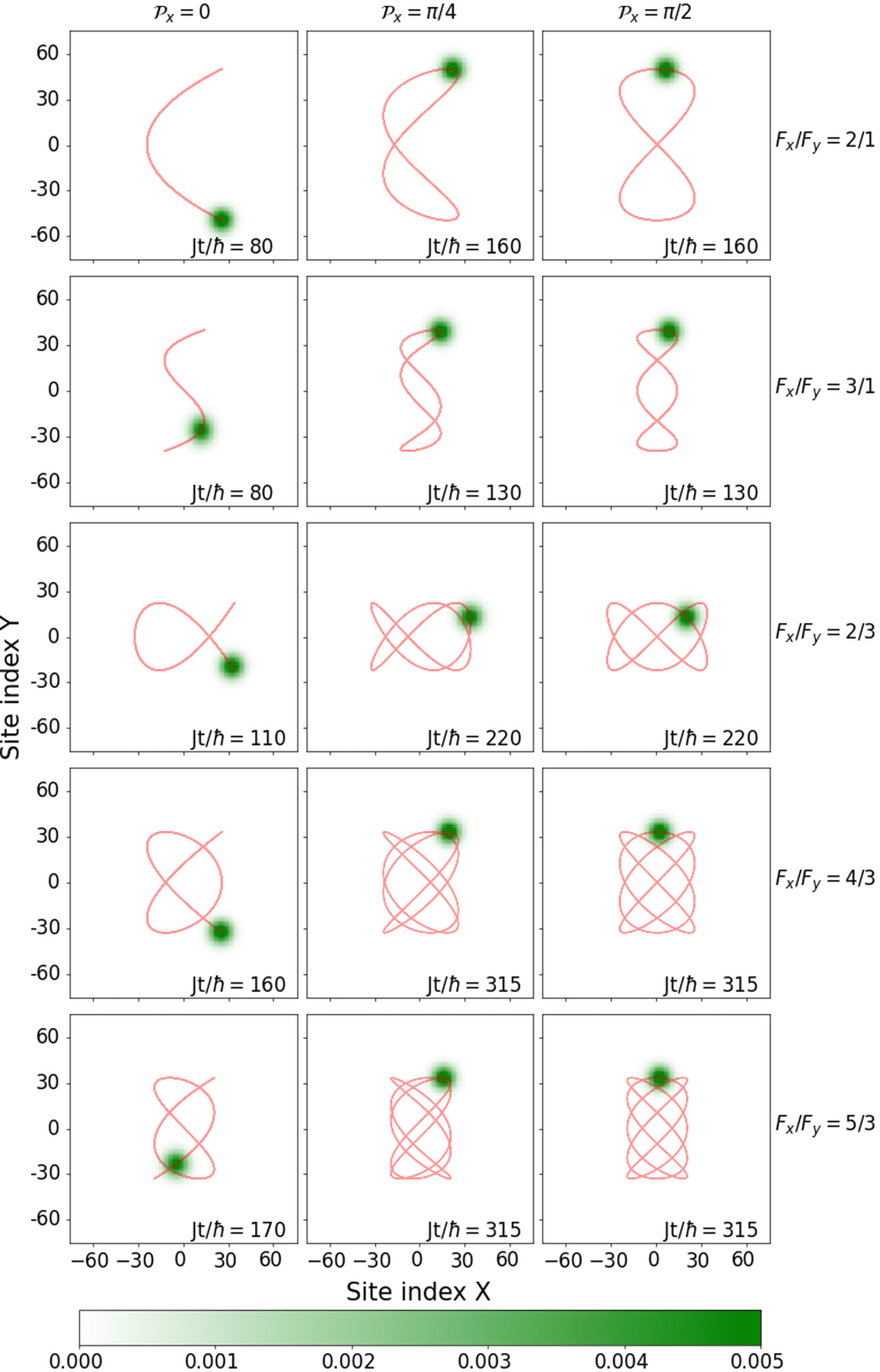}
\caption{Examples of the Lissajous trajectories traced by a Gaussian wave packet for selected lattice parameters $F_x/F_y$ and the initial momentum of the packet ${\cal P}_x$. Each plot shows the temporal density distribution at the selected moment (indicated for each plot separately) as well as the entire Lissajous curve. In all cases, the width of the wave packet is set to $\sigma=5$. Corresponding movies presenting a whole evolution are accessible online~\cite{2024JaczewskiSowinskiZenodo}.} \label{Fig5}
\end{figure}

It is straightforward to adjust lattice and initial state parameters to make the density distribution moving along different Lissajous curves. In Fig.~\ref{Fig5} we present several quantum mechanical scenarios in which the density distribution follows the most popular curves recognized in classical mechanics. Typically they are labeled by rational ratios of frequencies $\Omega_x/\Omega_y$ (controlled by the ratio $F_x/F_y$) and relative phase shifts $\varphi$ (controlled by initial momentum ${\cal P}_x$). Corresponding movies presenting a whole evolution are accessible online~\cite{2024JaczewskiSowinskiZenodo}.

\section{Conclusions}
In this work, we discussed the possibility of observing the manifestations of classical dynamics in the motion of a quantum particle on a two-dimensional tilted lattice. We originated from the well-known observation that in a periodic potential with a constant force, a quantum particle oscillates, and its probability density distribution does not change over time (for a sufficiently wide wave packet). We then pointed out that in the case of a two-dimensional lattice, the combination of two oscillatory motions can result in the motion of the packet along a Lissajous curve well known in classical mechanics. We have presented in detail the reasoning that allows us to predict the parameters for which the desired Lissajous curve can be obtained, and thus we have given a precise recipe for how to prepare the system to realize this.

The analysis presented here can be straightforwardly extended to the three-dimensional case, in which the wave packet will follow a three-dimensional Lissajous trajectory. One of the open questions we leave for further analysis concerns the stability of the presented solutions for wave packets formed by many interacting particles, for example, bosons with contact interactions or multi-component mixtures. Exploring quantum correlations induced by interactions during this kind of motion in systems of several particles (bosons as well as fermions) is also an interesting direction to extend previous results in one-dimensional systems~\cite{science1260364,PhysRevA.96.043629,PhysRevA.101.052341,PhysRevA.102.043326,SciRep22056,PhysRevLett.127.100406,PhysRevA.104.033306,PhysRevLett.129.050601,PhysRevA.109.043308,PhysRevB.110.014302}. We believe that presented analysis may also be helpful in better understanding of regular dynamics in the so-called Fock-state-lattices lacking transitional invariance~\cite{LarsonBook,2023Larson}.

This research was supported by the (Polish) National Science Centre within OPUS Project No. 2023/49/B/ST2/03744. For the purpose of Open Access, the authors have applied a CC-BY public copyright licence to any Author Accepted Manuscript version arising from this submission.

Data sets generated during the current study are available from the corresponding author on reasonable request.

\bibliography{biblio}

\end{document}